\documentclass[
reprint, amsmath,amssymb,aps,prl,superscriptaddress
]{revtex4-1}

\usepackage{graphicx}
\usepackage{amsmath} 
\usepackage[colorlinks,urlcolor=blue,linkcolor=blue,citecolor=blue]{hyperref}
\usepackage{amsfonts}
\usepackage{bbm}
\usepackage{braket}
\usepackage{caption}
\usepackage{graphicx}
\usepackage{float}
\usepackage{subcaption}
\usepackage{url}
\usepackage{qcircuit}
\usepackage{amssymb}
\usepackage{qcircuit}
\usepackage{caption}
\usepackage{siunitx}
\usepackage{braket}
\usepackage{multirow}
\usepackage{color}

\newcommand{\dqc}{Duke Quantum Center, Duke University, Durham, NC 27701, USA}

\begin{document}

\preprint{APS/123-QED}

\title{Crosstalk Suppression in Individually Addressed Two-Qubit Gates in a Trapped-Ion Quantum Computer}

\author{Chao Fang}
\email{chao.fang@duke.edu}
\affiliation{\dqc}
\affiliation{Department of Electrical and Computer Engineering, Duke University, Durham, NC 27708, USA}
\author{Ye Wang}
\email{ye.wang2@duke.edu}
\affiliation{\dqc}
\affiliation{Department of Electrical and Computer Engineering, Duke University, Durham, NC 27708, USA}
\author{Shilin Huang}
\affiliation{\dqc}
\affiliation{Department of Electrical and Computer Engineering, Duke University, Durham, NC 27708, USA}
\author{Kenneth R. Brown}
\affiliation{\dqc}
\affiliation{Department of Electrical and Computer Engineering, Duke University, Durham, NC 27708, USA}
\affiliation{Department of Physics, Duke University, Durham, NC 27708, USA}
\affiliation{Department of Chemistry, Duke University, Durham, NC 27708, USA}
\author{Jungsang Kim}
\email{jungsang.kim@duke.edu}
\affiliation{\dqc}
\affiliation{Department of Electrical and Computer Engineering, Duke University, Durham, NC 27708, USA}
\affiliation{Department of Physics, Duke University, Durham, NC 27708, USA}
\affiliation{IonQ, Inc., College Park, MD 20740, USA}

\date{\today}

\begin{abstract} 
Crosstalk between target and neighboring spectator qubits due to spillover of control signals represents a major error source limiting the fidelity of two-qubit entangling gates in quantum computers. We show that in our laser-driven trapped-ion system coherent crosstalk error can be modelled as residual $X\hat{\sigma}_{\phi}$ interaction and can be actively cancelled by single-qubit echoing pulses. We propose and demonstrate a crosstalk suppression scheme that eliminates all first-order crosstalk, yet only requires local control by driving rotations solely on the target qubits. We report a two-qubit Bell state fidelity of $99.52(6) \%$ with the echoing pulses applied after collective gates and $99.37(5) \%$ with the echoing pulses applied to each gate in a 5-ion chain. This scheme is widely applicable to other platforms with analogous interaction Hamiltonians.
\end{abstract}

\maketitle


Improving the performance of two-qubit entangling gates on a scalable physical platform is one of the key challenges in realizing a practical quantum computer. 
High two-qubit gate fidelities reaching the requirements of fault-tolerant quantum computation~\cite{fowler2012surface, knill2005quantum} have been realized in limited two-qubit systems across various qubit platforms \cite{sung2021high, stehlik2021high, madjarov2020, noiri2022, xue2022}, with the highest reported fidelities using trapped-ion qubits \cite{ballance2016high, gaebler2016high, clark2021high, srinivas2021high}. When extending high-fidelity two-qubit gates to a large array of qubits, crosstalk (also known as addressing error) needs to be considered, which is caused by unwanted spillover of the control signal onto neighboring spectator qubits when addressing the target qubits. It represents a major error source in noisy intermediate-scale quantum (NISQ) systems and can potentially break error correcting codes used in fault-tolerant quantum computation \cite{merrill2014, debroy2020}.

The M\o lmer-S\o rensen (MS) gate is a widely-used two-qubit gate protocol in trapped-ion quantum computers, where qubits are entangled through Coulomb-coupled collective motion of the ions \cite{sorensen1999quantum}. MS gates have been demonstrated with fidelity $>99.9\%$ utilizing axial motional modes \cite{ballance2016high, gaebler2016high} and $99.5\%$ utilizing radial motional modes with individual addressing \cite{wang2020}. Detailed models of the various MS gate error sources have been developed, but crosstalk errors are often neglected in error analysis limited to a two-ion crystal \cite{ballance2014thesis}, or assumed to be insignificant with the addressing laser beams having ideal Gaussian profiles \cite{wu2018}. In trapped-ion systems, crosstalk mainly occurs as a result of residual illumination of neighboring ions by laser light during gate applications due to the finite size of the focused beam spot. On the hardware level, crosstalk can be reduced through careful design of optical addressing systems that minimize optical aberrations \cite{crain2014, shih2021}, or spatial modulation of multiple beams that interfere at qubit locations to achieve super-resolution addressing \cite{saffman2004, shen2013}. On the level of error correcting codes and circuits, the effect of crosstalk errors can be mitigated by choosing the optimal code and qubit arrangement in the circuitry \cite{debroy2020}. As crosstalk errors are mostly coherent in nature, they can be effectively suppressed for single-qubit gates using spin-echo methods \cite{nigg2014}, composite pulse sequences \cite{brown2004sk1, MerrillPRA2014} or dynamical decoupling techniques \cite{viola1999}. For two-qubit entangling gates, a gate-level crosstalk suppression method has been proposed where echoing pulses are applied to all the neighboring spectator qubits to cancel the crosstalk interaction \cite{parrado2021}. While this scheme is shown to theoretically reduce the effect of crosstalk on the performance of the 7-qubit color code by four orders of magnitude, its implementation in a large scale system considerably increases the number of single-qubit gates.
 
Here we present a novel scheme to suppress first-order crosstalk on the gate level by applying echoing pulses only to the target qubits. We model the coherent crosstalk error as residual entanglement between the spectator qubits and the target qubits in the MS gate framework, and show that first-order crosstalk to all the spectator qubits can be cancelled by our echoing scheme. We implement both echoing techniques experimentally and demonstrate high-fidelity two-qubit gates ($99.5 \%$) in the presence of considerable crosstalk in a 5-ion chain.

For the crosstalk error model in this work, we focus on the MS gate in the $XX$-basis, which is described by the unitary $XX(\theta) = \exp(-i\theta XX)$, where $\theta$ is the geometric phase of the MS gate. In a linear ion chain system, we consider ion 1 and ion 2 to be the target qubits which are addressed by two tightly focused beams, and the rest of the ions to be the spectators, as shown in Fig. 1(a). Imperfect optical addressing leads to gate crosstalk between a target ion $i$ and a spectator ion $j$, which is defined as the ratio of Rabi frequencies, $\epsilon_{ij} = \Omega_j/\Omega_i$ $(i = 1,2)$, when resonantly driving single-qubit gates on ion $i$. In state-of-the-art trapped-ion experiments, $\epsilon_{ij}$ is in the range of $1 \sim 3 \%$ for nearest neighbors \cite{debnath2016, pogorelov2021, egan2021thesis} and can be below $1 \%$ with high-performance optical addressing technologies \cite{wang2020, pogorelov2021}. The leading order effect of crosstalk is an additional MS interaction $X^{(i)}\hat{\sigma}_{\phi}^{(j)}$ where $\hat{\sigma}_{\phi}^{(j)} = \cos(\phi)X^{(j)} + \sin(\phi)Y^{(j)}$, as shown in Fig. 1(b). The ideal MS gate unitary operator is then replaced with 
\begin{align}\label{eq:U}
    U_{\text{xtalk}}(\theta) = &\exp\big(-i{\theta}X^{(1)}X^{(2)} \big) \exp\big[-i\sum\limits_{j\ne 1,2}(\theta_{1,j}X^{(1)}\hat{\sigma}_{\phi_j}^{(j)} \nonumber\\
    &\quad+ \theta_{2,j}X^{(2)}\hat{\sigma}_{\phi_j}^{(j)}) \big],
\end{align}
where $j$ is the index of all the affected spectator ions, and $\theta_{1,j}$ and $\theta_{2,j}$ are the respective geometric phases of the MS interactions between each target ion and ion $j$. For arbitrary spectator qubit states, this can generate unwanted entanglement between spectator qubits and data qubits. Entanglement monogamy dictates that two maximally entangled parties cannot share entanglement with a third party \cite{coffman2000}. Therefore, this unwanted residual entanglement with spectator qubits reduces the fidelity of entangling gates on the target qubits. If we consider only one spectator ion, e.g. $j = 3$, for simplicity, and initiate the qubit states of the two target and the spectator ions to $\ket{0}$, after applying the maximally entangling gate $XX(\pi/4)$ to the target ions, the fidelity of the target ion Bell state as well as the unwanted excitation of the population of the spectator ion can be calculated using Eq.~\ref{eq:U} and are given by
\begin{equation}\label{eq:F}
    \mathcal{F}_{\text{xtalk}} = \frac{1}{4}\big(1 + \cos(\theta_{1,3}) \big)\big(1 + \cos(\theta_{2,3}) \big),
\end{equation}
\begin{equation}\label{eq:P}
    P_{\text{ion3}} = \frac{1}{2}\big(1 - \cos(\theta_{1,3})\cos(\theta_{2,3}) \big),
\end{equation}
where $P_{\text{ion3}}$ is equal to the $\ket{01}$ and $\ket{10}$ population of the target ions.

\begin{figure}[!htb]
\center{\includegraphics[width=0.48 \textwidth]{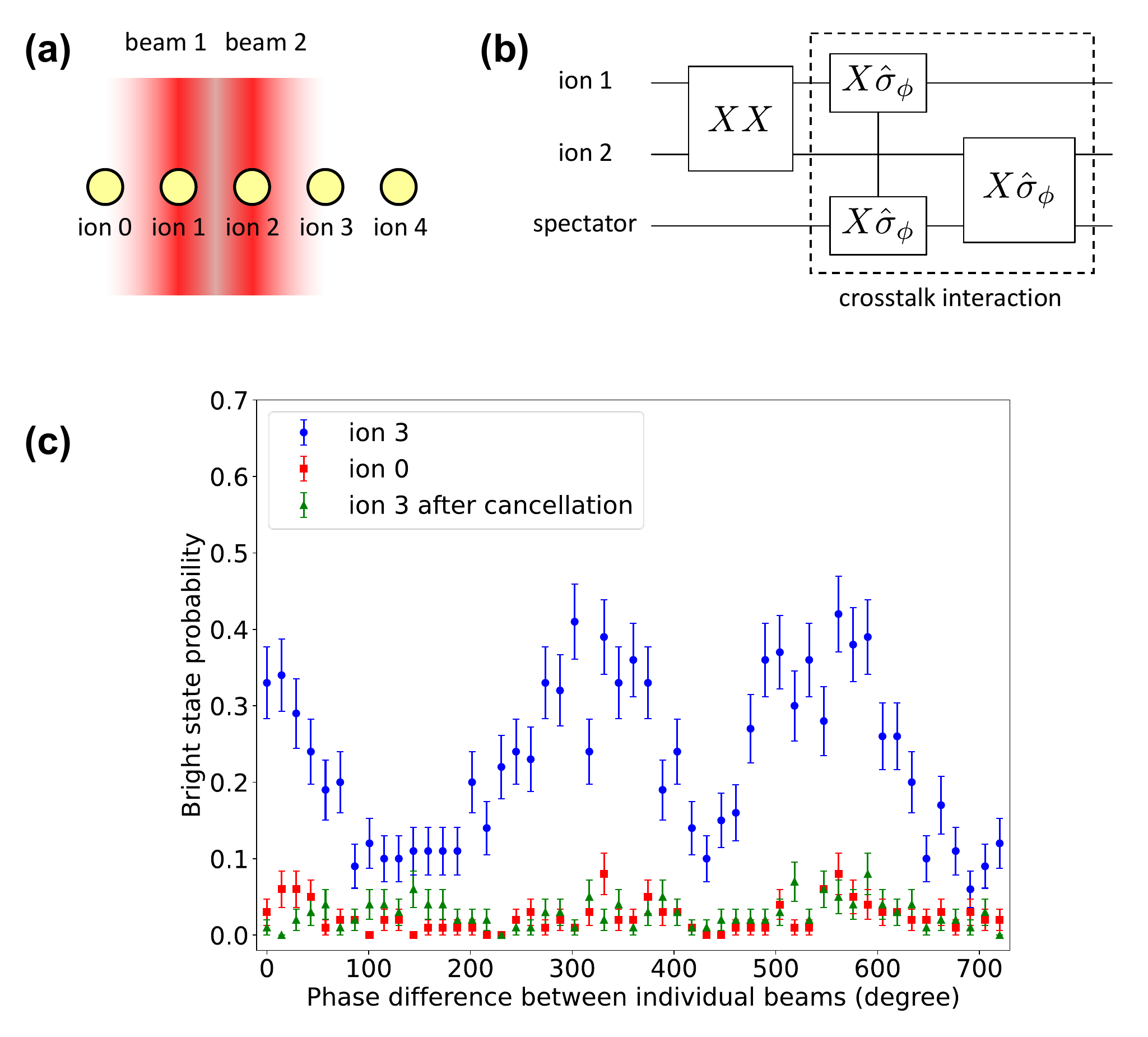}}
\caption{\label{Crosstalk illustration} \textbf{(a)} Schematic of a 5-ion chain with two tightly focused beams addressing ion 1 and ion 2, and the rest of the ions considered to be the spectators impacted by intensity crosstalk. \textbf{(b)} Circuit model of the effect of crosstalk with one spectator qubit. The residual entanglement between each target qubit and the spectator qubit is expressed as the MS interaction $X\hat{\sigma}_{\phi}$. \textbf{(c)} Populations of ion 0 and ion 3 (red and blue) after applying 21 consecutive $XX(\pi/4)$ to ion 1 and ion 2. The population of ion 3 varies as the effective crosstalk depends on $\phi_{\text{beam}}$. The population of ion 0 sees a much smaller excitation since ion 0's coupling to the motional modes mostly involved in the MS gate is weaker. The population of ion 3 after crosstalk suppression using the echoing technique is also shown (green). The average population drops from 0.25 to 0.03.}
\end{figure}

The error parameters $\theta_{1,j}$ and $\theta_{2,j}$ as well as the angle $\phi_j$ in Eq.~\ref{eq:U} depend on the effective Hamiltonian of ion $j$ which is a product of the two individual addressing beams interfering at the location of ion $j$. This Hamiltonian is a function of not only the gate crosstalk $\epsilon_{1j}$ and $\epsilon_{2j}$, but also the optical phase difference $\phi_{\text{beam}}$ between the two addressing beams (see supplementary material). In our system, the two addressing beams are delivered via beam paths that are spatially separated. As a result, this phase difference $\phi_{\text{beam}}$ drifts slowly over a timescale characterized by the stability of the Raman beam paths ($\sim 100$s of ms), giving rise to uncertainty in effective crosstalk errors. 

Fig. 1(c) shows the measured populations of two spectators, ion 0 and ion 3 in the chain configuration shown in Fig. 1(a), after applying 21 consecutive $XX(\pi/4)$ gates to ion 1 and ion 2. The optical phase of one addressing beam is scanned from 0 through $4\pi$ while the phase of the other beam is set to 0. In the experiment, the gates are implemented by driving stimulated Raman transitions using a picosecond pulsed laser and discrete frequency modulation (FM) is used to minimize the error from coupling to all collective motional modes during the MS gate \cite{leung2018robust, leung2018arbitrary, kang2021}. Details of the experimental setup are described in Ref. \cite{wang2020}. As shown in Fig. 1(c), the population of ion 3 varies within a large range as a function of the applied phase difference. The irregular oscillation (more than 2.5 cycles) indicates that the actual $\phi_{\text{beam}}$ is drifting over the duration of taking the experimental data ($\sim 30$ sec). The population of ion 0 sees a much smaller excitation than ion 3 despite it is subjected to similar crosstalk, since ion 0 has weaker coupling to the motional modes mostly involved in the MS gate and the motional mode participation by the spectator ion $j$ also contributes to $\theta_{1,j}$ and $\theta_{2,j}$.

Due to the coherent nature of crosstalk, as shown in Eq.~\ref{eq:U}, this error can be actively cancelled by applying single-qubit spin-echo pulses in the middle of the gate(s), reversing the crosstalk interaction during the second half of the MS evolution. The echoing scheme proposed in Ref. \cite{parrado2021} (which we call ``neighbor suppression'') is illustrated in the circuit of Fig. 2(a). The scheme's working principle follows a simple proof using Eq.~\ref{eq:U}, $\left[\prod\limits_{j\ne 1,2}Z^{(j)}\emph{U}_{\text{xtalk}}(\frac{\theta}{2}) \right]^2 = \exp\big(-i\theta X^{(1)}X^{(2)} \big)$. A single $XX(\pi/4)$ is added after the second $Z(\pi)$ pulse in order to generate the Bell state whose fidelity can be characterized by quantum state tomography. It should be noted that since $\phi_j$ is not well-defined due to drifting $\phi_{\text{beam}}$, we must use $Z$ gate to reverse the $X\hat{\sigma}_{\phi_j}$ interaction rather than $Y$ gate. We experimentally verify the crosstalk suppression scheme by steering one of the addressing beams to the spectator ion to drive $Z(\pi)$ rotations (implemented as $X(\pi)Y(\pi)$) using micro-electromechanical system (MEMS) mirrors \cite{crain2014}. The echoing pulses are only applied to ion 3 since the crosstalk error in this case is dominated by $\theta_{1,3}$ and $\theta_{2,3}$ as shown in Fig. 1(c). The green triangles in Fig. 1(c) show the population of ion 3 after applying 21 $XX(\pi/4)$ gates with the echoing scheme, averaging 0.03 in contrast to 0.25 without crosstalk suppression.

This neighbor suppression method requires applying single-qubit gates to all the affected spectator qubits, which can be up to 8 ions including the nearest and next-nearest neighbors in a long-chain system, and more if farther spectator ions have non-negligible crosstalk \cite{egan2021thesis}. We propose a scheme that achieves cancellation of all first-order crosstalk by rotating only the two target qubits, significantly reducing the experimental resource overhead (named ``local suppression''). Even with only one spectator ion involved in our experiment, this method reduces the control overhead as it bypasses use of MEMS beam-steering. Fig. 2(b) shows the circuit for implementing local suppression scheme, where either $Y(\pi)$ or $Z(\pi)$ can be used as the echoing pulses on the target qubits. Since $Y^{(1)} \otimes I^{(j)}$ anti-commutes with $X^{(1)}\hat{\sigma}_{\phi_j}^{(j)}$, $Y^{(2)} \otimes I^{(j)}$ anti-commutes with $X^{(2)}\hat{\sigma}_{\phi_j}^{(j)}$ and $Y^{(1)} \otimes Y^{(2)}$ commutes with $X^{(1)}X^{(2)}$, it can be easily shown from Eq.~\ref{eq:U} that $\big[Y^{(1)} \otimes Y^{(2)}\emph{U}_{\text{xtalk}}(\frac{\theta}{2}) \big]^2 = \exp\big(-i{\theta}X^{(1)}X^{(2)} \big)$. We note that the efficacy of both neighbor and local suppression schemes is not affected by the slow drift of $\phi_{\text{beam}}$, as the drift is negligible on the timescale required to complete the crosstalk cancellation.

\begin{figure}[!htb]
(a)\centerline{
\Qcircuit @C=0.5em @R=0.6em {
& \lstick{\ket{0}} & \multigate{1}{XX(\frac{\pi}{4})^{\times n}} & \qw & \multigate{1}{XX(\frac{\pi}{4})^{\times n}} & \qw & \multigate{1}{XX(\frac{\pi}{4})} & \meter \\
& \lstick{\ket{0}} & \ghost{XX(\frac{\pi}{4})^{\times n}} & \qw & \ghost{XX(\frac{\pi}{4})^{\times n}} & \qw & \ghost{XX(\frac{\pi}{4})} & \meter \\
& \lstick{\ket{0}} & \qw & \gate{Z} & \qw & \gate{Z} & \qw & \meter
}}\\~\\~\\
(b)\centerline{
\Qcircuit @C=0.5em @R=0.6em {
& \lstick{\ket{0}} & \multigate{1}{XX(\frac{\pi}{4})^{\times n}} & \gate{Y} & \multigate{1}{XX(\frac{\pi}{4})^{\times n}} & \gate{Y} & \multigate{1}{XX(\frac{\pi}{4})} & \meter \\
& \lstick{\ket{0}} & \ghost{XX(\frac{\pi}{4})^{\times n}} & \gate{Y} & \ghost{XX(\frac{\pi}{4})^{\times n}} & \gate{Y} & \ghost{XX(\frac{\pi}{4})} & \meter \\
& \lstick{\ket{0}} & \qw & \qw & \qw & \qw & \qw & \meter
}}\\~\\~\\
(c)\centerline{
\Qcircuit @C=0.8em @R=0.8em {
&&& \mbox{repeat N times} &&& \\
& \lstick{\ket{0}} & \multigate{1}{XX(\frac{\pi}{8})} & \gate{Y} & \multigate{1}{XX(\frac{\pi}{8})} & \gate{Y} & \meter \\
& \lstick{\ket{0}} & \ghost{XX(\frac{\pi}{8})} & \gate{Y} & \ghost{XX(\frac{\pi}{8})} & \gate{Y} & \meter \\
& \lstick{\ket{0}} & \qw & \qw & \qw & \qw & \meter \gategroup{2}{3}{3}{6}{0.6em}{--}
}}
\caption{\label{Circuits}Circuit diagrams for crosstalk suppression in the MS gate. \textbf{(a)} Neighbor suppression performed in the collective-gate echoing experiments where $Z(\pi)$ rotations are driven on the spectator ion. A single $XX(\pi/4)$ is added after the second $Z(\pi)$ pulse to generate the Bell state for fidelity measurement. \textbf{(b)} Local suppression performed in the collective-gate echoing experiments where $Y(\pi)$ pulses are applied to the target ions. \textbf{(c)} Local suppression performed in the individual-gate echoing experiments where the $XX(\pi/4)$ is split into two half MS evolution and the echoing $Y(\pi)$ pulses are applied after each $XX(\pi/8)$.}
\end{figure}
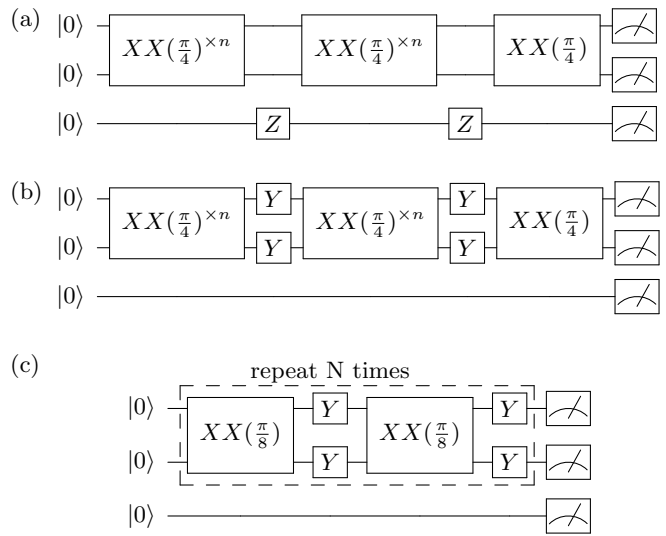

Two sets of experiments are conducted to showcase the effectiveness of the echoing technique in improving the fidelity of the two-qubit MS gate in a 5-ion chain. In the first set of experiments, we apply a sequence of 1, 9, 13 and 21 $XX(\pi/4)$ to ion 1 and ion 2 in the chain configuration shown in Fig. 1(a) and then measure the Bell state fidelity from the final state $\ket{01}$ and $\ket{10}$ population and parity contrast \cite{leibfried2003experimental}. We implement both crosstalk suppression schemes in the collective-gate echo fashion as illustrated in the circuits of Fig. 2(a) and (b) and compare the measured fidelity to that of the native gate without crosstalk suppression. For neighbor suppression, the $Z(\pi)$ rotations are only applied to ion 3. The SK1 composite pulse sequence is used to implement the single-qubit gates to eliminate crosstalk from the echoing pulses themselves \cite{brown2004sk1}. The Bell state infidelity along with the spectator ion population (sum of ion 0 and ion 3 populations) after the native gates are plotted in Fig. 3(a). The infidelity and the spectator ion population without crosstalk suppression do not follow the typical quadratic trend for accumulated coherent errors due to the uncertainty in our crosstalk errors. The range of possible state infidelity is numerically simulated by assuming a randomly drifting $\phi_{\text{beam}}$ that remains constant for the duration of each gate (shown by the light-blue shaded area, see supplementary material for details of the crosstalk error simulation). With crosstalk suppression, the infidelity is dominated by stochastic errors which accumulate in a linear fashion with the number of gates. Using a linear fit for the data, we extract a two-qubit Bell state fidelity of $99.33(6) \%$ with neighbor suppression and $99.52(6) \%$ with local suppression. The lower fidelity using neighbor suppression is attributed to the small residual excitation of ion 0 population, which is not actively cancelled by the scheme. The remaining $0.48\%$ error is consistent with the simulated error budget based on the noise parameters in our system, which is dominated by motional dephasing (see Ref. \cite{wang2020} and supplementary material for details of the simulated error budget).

While the collective-gate echo approach shown in Fig. 2(a) and (b) is useful for characterizing the two-qubit gate fidelity with suppressed crosstalk errors, we want to apply the maximally entangling gate as a single $XX(\pi/4)$ and implement crosstalk suppression on the individual-gate level in most practical circuits. In the second set of experiments, the $XX(\pi/4)$ is split into two half MS evolution and the echoing $Y(\pi)$ pulses are applied after each $XX(\pi/8)$, as shown in Fig. 2(c) (``individual-gate echo'' experiments). We apply a sequence of 1, 9, 13, 17 and 21 gates to ion 1 and ion 3 using local suppression scheme. For the native gates, only ion 2 (center ion) population is affected by crosstalk with negligible excitation of ion 0 and ion 4 populations due to their low motional mode participation. The results with and without crosstalk suppression are compared in Fig. 3(b). A linear fit for the data using local suppression produces a two-qubit 
Bell state fidelity of $99.37(5) \%$. The fidelity is lower than that using collective-gate local suppression as four single-qubit gates are added to each $XX(\pi/4)$. The fidelity of the SK1 single-qubit gates in our system has been characterized using gate set tomography (GST) analysis \cite{blume2017} to be about $99.93\%$ for $Y(\pi/2)$ \cite{zhang2022}. Simulation including the single-qubit gate fidelity penalty agrees with the $0.63\%$ residual error.

\begin{figure}[!htb]
\center{\includegraphics[width=0.48 \textwidth]{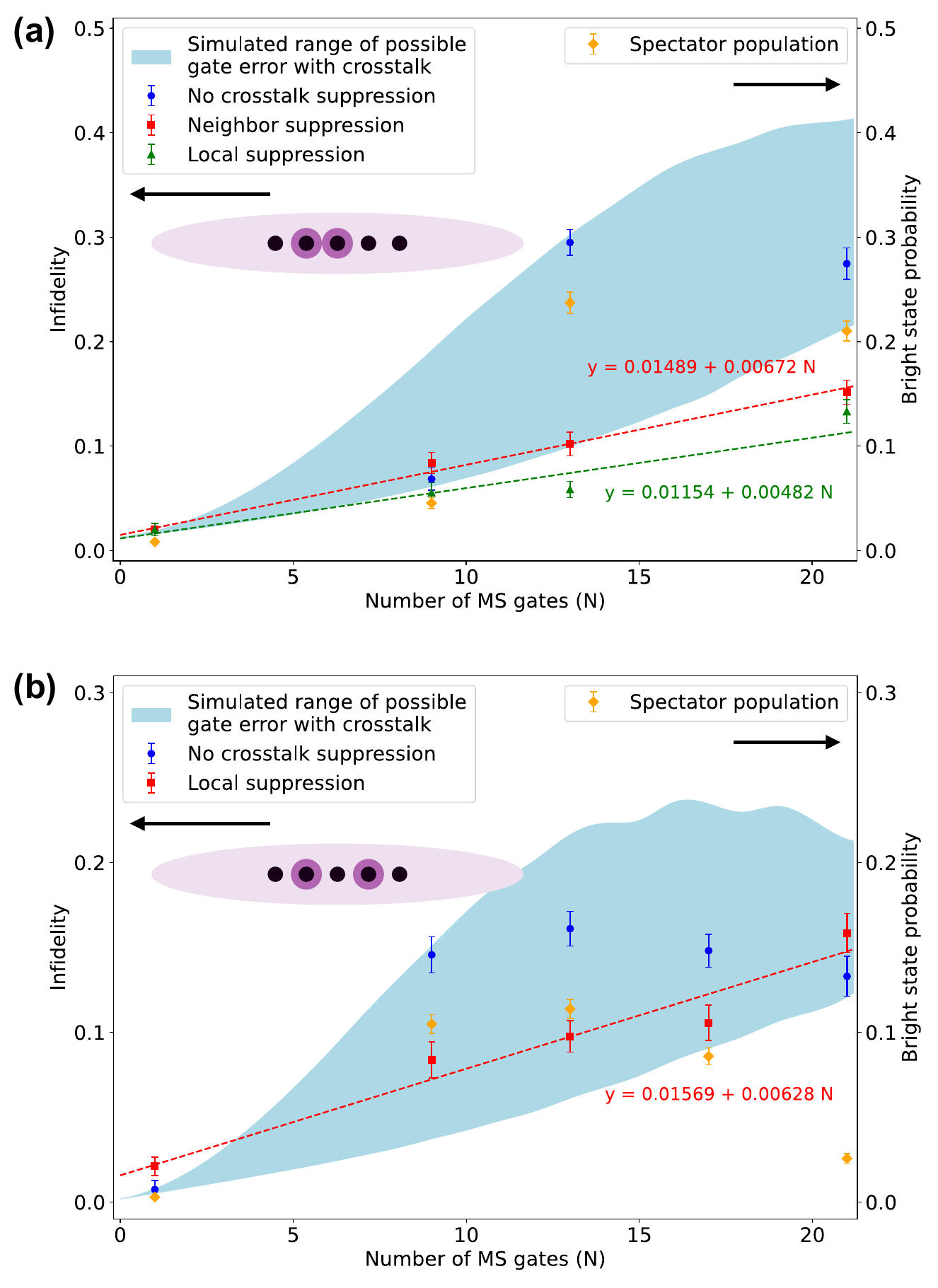}}
\caption{\label{fig:fidelity} Results for the two-qubit Bell state infidelity and spectator ion population plotted against the number of consecutive gates for \textbf{(a)} the collective-gate echo experiments comparing the gates without crosstalk suppression to the gates with neighbor/local suppression and \textbf{(b)} the individual-gate echo experiments comparing the gates without crosstalk suppression to the gates with local suppression. The simulated gate error range is shown by the shaded area. The infidelity contribution of a single MS gate with crosstalk suppression is post-processed from linear fit.
}
\end{figure}

For the crosstalk error simulation, instead of simply assuming the upper bound of the error range to be when the two addressing beams always constructively interfere ($\phi_{\text{beam}} = 0$) and the lower bound to be when they always destructively interfere ($\phi_{\text{beam}} = \pi$), which is only possible if $\phi_{\text{beam}}$ stays constant through the duration of the entire experiment, we take into account the effects of 1) the slow drift of $\phi_{\text{beam}}$, and 2) the continuous phase shift in $\phi_j$ arising from light shift due to AC Stark effect caused by the MS gate, to more realistically model the crosstalk error behavior. The first effect is simulated by a 1D random walk of $\phi_{\text{beam}}$ after each gate, accumulating for 100 repeated applications of consecutive gates. The second effect is modelled by adding a constant shift to $\phi_j$ in Eq.~\ref{eq:U} after each gate, which accumulates with the number of gates but not with repeated experiments. This phase shift is the result of the small difference in qubit frequencies between the target ions and the spectator ions, as the addressing laser beams induce much larger light shift on the target ions \cite{mizrahi2014, lee2016}. We calibrate the light shift to be $\sim 4^{\circ}$ per MS gate in the collective-gate echo experiments and $\sim 6^{\circ}$ per MS gate in the individual-gate echo experiments.

As shown in Fig. 3, with the combination of both effects, the upper bound of the simulated error range plateaus and the error range starts to taper as the number of gates increases, because the effect of crosstalk tends to partially cancel out for long gate sequences. The maximum step size for the random walk simulation of $\phi_{\text{beam}}$ can be estimated by forcing the simulated range to encompass all the measured infidelity data (blue circles), which is approximately a 2-Hz drift for both sets of experiments. This is consistent with the timescale over which our Raman beam paths are interferometrically stable, measured by Ramsey interference experiment (see supplementary material). We note that the dependence of effective crosstalk on the Raman beam path stability are insignificant for experimental systems with mostly shared beam paths for addressing beams \cite{debnath2016, egan2021}, and the crosstalk suppression scheme introduced in this work all but eliminates the dependence on the drift of our Raman beam paths.

In conclusion, the echoing pulse technique is a simple and effective way to suppress crosstalk errors in the two-qubit entangling gate thanks to the coherent nature of crosstalk between the target and spectator qubits. We report a crosstalk suppression scheme that only utilizes local control over the target qubits to cancel all crosstalk to the first order, improving upon the previously proposed method by significantly reducing the resource overhead required for applying gates to all the affected spectator qubits. Although the error model and the proof-of-principle experiments discussed in this work are limited to the MS gate, we note that the scheme can be applied to two-qubit gates based on light shift \cite{leibfried2003experimental} as well as other physical platforms with interaction Hamiltonians similar to trapped-ion qubits.

\acknowledgements
We thank Bichen Zhang for his contribution to the experiments and Zhubing Jia for helpful discussions. This work was primarily supported by the Office of the Director of National Intelligence - Intelligence Advanced Research Projects Activity through ARO contract W911NF-16- 1-0082 (experimental implementation and error simulation), by DOE BES award DE-SC0019449 (gate simulation), by National Science Foundation Expeditions in Computing award 1730104 (theoretical conception), and by DOE ASCR award DE-SC0019294 (theoretical conception).

\bibliographystyle{apsrev4-1} 
\bibliography{./References}

\clearpage

\onecolumngrid
\section{Supplementary Material}

\subsection{Gate Crosstalk in 5-Ion Chain Experiments}

Table I and II show the measured gate crosstalk $\epsilon_{ij}$ relevant to the two sets of experiments presented in this work, i.e. the collective-gate echo experiments and the individual-gate echo experiments. The crosstalk values are used in the simulation of crosstalk errors for the respective sets of experiments.

\begin{table}[!htb]
\setlength{\tabcolsep}{8pt}
\renewcommand{\arraystretch}{1.5}
\centering
\begin{tabular}{c|c  c  c  c  c}
    \hline
    \multirow{2}{0pt}{$i$} & \multicolumn{5}{c}{$j$} \\
    & ion 0 & ion 1 & ion 2 & ion 3 & ion 4 \\
    \hline
    beam 1 & $3.3 \%$ & NA & $4 \%$ & $1.6 \%$ & $<0.5 \%$ \\
    beam 2 & $3.6 \%$ & $2 \%$ & NA & $3.5 \%$ & $<0.5 \%$ \\
    \hline
  \end{tabular}
\caption{Crosstalk $\epsilon_{ij}$ for the collective-gate echo experiments. The level of nearest neighbor and next-nearest neighbor crosstalk is higher than what is typically seen in state-of-the-art trapped-ion experiments due to aberrations caused by contaminated component in the optical addressing system but still within the range of believable experimental conditions.
}
\end{table}

\begin{table}[!htb]
\setlength{\tabcolsep}{8pt}
\renewcommand{\arraystretch}{1.5}
\centering
\begin{tabular}{c|c  c  c  c  c}
    \hline
    \multirow{2}{0pt}{$i$} & \multicolumn{5}{c}{$j$} \\
    & ion 0 & ion 1 & ion 2 & ion 3 & ion 4 \\
    \hline
    beam 1 & $2.6 \%$ & NA & $2.4 \%$ & $0.5 \%$ & $<0.1 \%$ \\
    beam 2 & $<0.2 \%$ & $1.3 \%$ & $1.9 \%$ & NA & $1.6 \%$ \\
    \hline
  \end{tabular}
\caption{Crosstalk $\epsilon_{ij}$ for the individual-gate echo experiments. The crosstalk level is lower after fixing the contaminated component.
}
\end{table}

\subsection{Crosstalk Error Simulation}

The effective Hamiltonian of an ion crystal with target ions 1 and 2 addressed by two laser beams and crosstalk on the other ions is given by
\begin{align*}
H = \sum_{j,k} \frac{\eta_{j,k} \Omega_j}{2} \hat{\sigma}_{\phi_j}^{(j)}\left(a_k^{\dag}e^{i\varphi_k(t)} + a_k e^{-i \varphi_k(t)}\right),
\end{align*}
where $\eta_{j,k}$ is the coupling strength between ion $j$ and motional mode $k$, $\Omega_j$ is the effective carrier Rabi frequency of ion $j$, $\phi_j$ is the spin phase of ion $j$, and $\varphi_k(t)$ is the time-dependent motional phase of mode $k$ controlled by frequency modulation (FM). 
For the two target ions, $\phi_1$ and $\phi_2$ are the optical phases of the two addressing beams. By choosing a convenient laser frame, we let $\phi_1 = 0$ and $\phi_2 = \phi_{\text{beam}}$.
For the spectator ions $j \ne 1,2$, $\Omega_j$ and $\phi_j$ are defined by the equation $$\Omega_j\hat{\sigma}_{\phi_j}^{(j)} = \epsilon_{1j}\Omega_1\hat{\sigma}_{\phi_1}^{(j)}+\epsilon_{2j}\Omega_2\hat{\sigma}_{\phi_2}^{(j)}.$$

Given an FM pulse sequence of duration $\tau$, the residual entanglement between ion $j$ and mode $k$ is determined by a displacement~\cite{leung2018robust}
\begin{align*}
    \alpha_j^k(\tau) = \frac{\eta_{j,k}\Omega_j}{2} \int_{0}^\tau e^{i\varphi_k(t)} dt.
\end{align*}
Our FM pulse sequence satisfies that $\alpha_j^k(\tau) = 0$ for all the modes. In the absence of noise, all ions are decoupled from the motional modes and the MS unitary action is $$U = \exp\left(-i\sum_{j_1<j_2} \theta_{j_1,j_2} \hat{\sigma}_{\phi_{j_1}}^{(j_1)} \hat{\sigma}_{\phi_{j_2}}^{(j_2)}\right),$$
where
\begin{align*}
    \theta_{j_1,j_2} = \frac{\Omega_{j_1}\Omega_{j_2}}{2} \sum_{k} \eta_{j_1,k}\eta_{j_2,k}\int_{0}^{\tau} dt \int_{0}^{t} dt' \sin\left(\varphi_k(t)-\varphi_k(t')\right).
\end{align*}
As the second-order rotation angles $\theta_{j_1,j_2}$ for $j_1,j_2\ne 1,2$ are negligible, Eq.~\ref{eq:U} in the main text can be obtained by choosing the appropriate qubit reference frame.

To efficiently estimate the effect of both coherent crosstalk error and stochstic errors with concatenated gates and 100 repeated experiments, 
instead of adopting a direct master-equation simulation, 
we assume that the contributions of infidelity from coherent and stochastic errors are additive. The infidelity caused by coherent crosstalk error can be analytically calculated given $\Omega_{1}$, $\Omega_{2}$, $\phi_{\textrm{beam}}$, $\epsilon_{1j}$ and $\epsilon_{2j}$.
We further assume that the infidelity caused by stochastic errors scales linearly with the number of gates. 
The infidelity caused by stochastic errors for each gate is obtained by numerically solving the master equation using the same FM pulse sequence that is optimized for the experiment. Details of the numerical simulation are described in Ref.~\cite{wang2020}.

\subsection{Error Budget Simulation}

With crosstalk suppression, the remaining stochastic errors in our MS gates are analyzed using numerical simulation. We consider the three most dominant error sources: motional dephasing, motional heating and laser (spin) dephasing. Table III shows their contributions to the error budget for the $99.52(6) \%$ fidelity two-qubit gate demonstrated in the collective-gate echo experiments using local suppression. Other error sources such as spontaneous emission and imperfect FM solution have a sum of contributions on the order of $10^{-4}$.

\begin{table}[!htb]
\setlength{\tabcolsep}{8pt}
\renewcommand{\arraystretch}{1.5}
\centering
\begin{tabular}{c|c}
\hline
Error source & Simulated error ($10^{-3}$) \\
\hline
Motional dephasing & $3.8 \pm 0.5$ \\
Motional heating & $0.42 \pm 0.02$ \\
Laser dephasing & $0.36 \pm 0.01$ \\
Total & $4.6 \pm 0.5$ \\
\hline
Experiment total & $4.8 \pm 0.6$ \\
\hline
\end{tabular}
\caption{Simulated error budget and experimental measurement of the remaining MS gate error in the collective-gate echo experiments using local suppression. Three major sources of stochastic errors are included in the error budget simulation, with motional dephasing dominating the total error in our experiments.
}
\end{table}

The effect of all three error sources are simulated by solving the master equation in Lindblad form, following the method described in the supplementary material of Ref. \cite{wang2020}. We measure the motional and laser coherence times in our experimental system to be 8(1) ms and 494(18) ms (see the next section of supplementary material), respectively. The motional heating rate in our ion trap is characterized to be 614(18) quanta/s for the center-of-mass mode and $<10$ quanta/s for the other modes in a 5-ion chain.

\subsection{Ramsey Measurement of Raman Beam Path Stability}
\begin{figure}[!htb]
\center{\includegraphics[width=0.48 \textwidth]{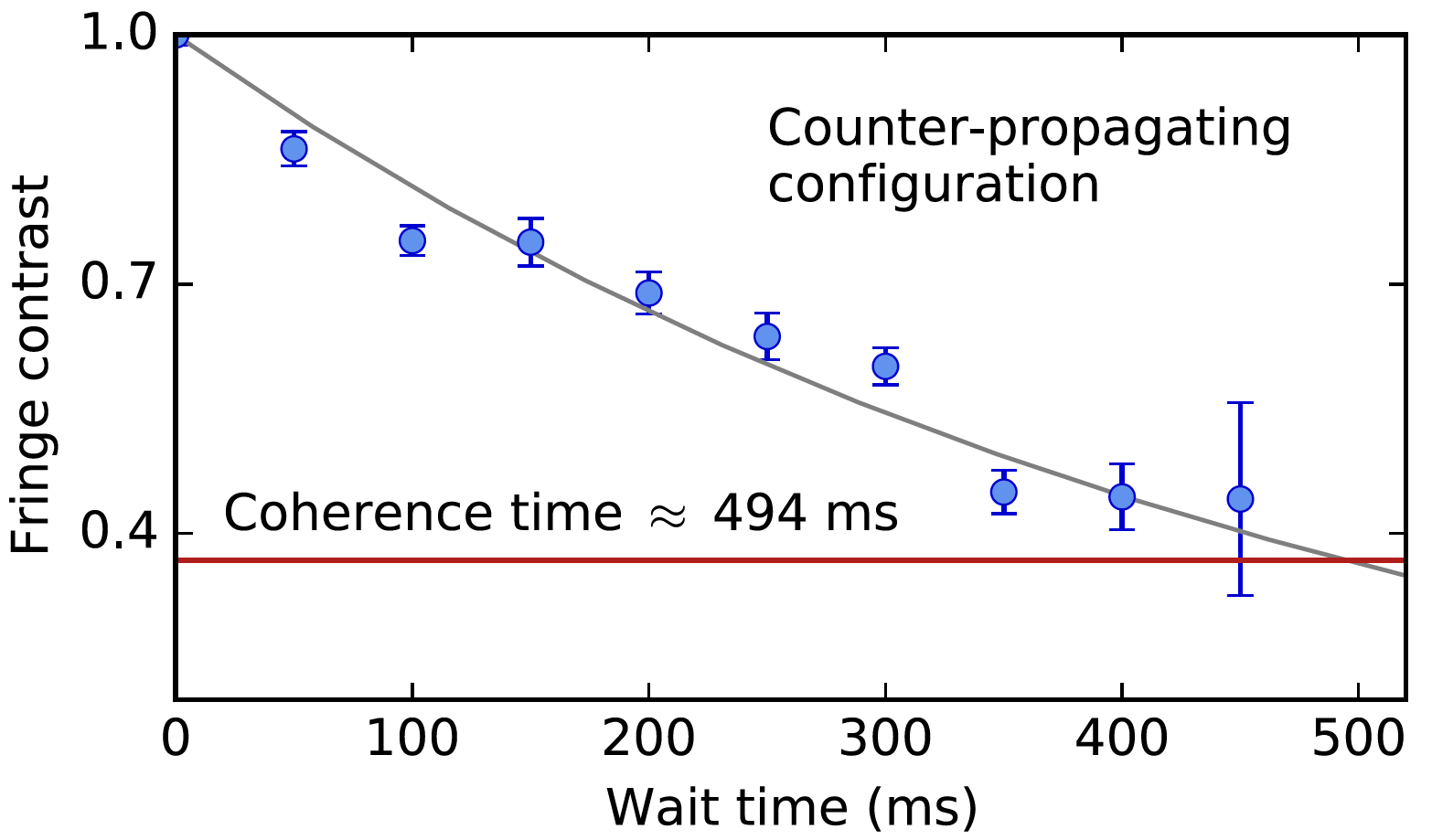}}
\caption{Ramsey interference experiment using two Raman beams in the counter-propagating configuration to characterize the stability of Raman beam paths. The Ramsey fringe contrast as a function of the wait time is fitted to exponential decay to extract a laser coherence time of 494(18) ms.
}
\end{figure}
The slow drift of $\phi_{\text{beam}}$ in our experimental system is characterized by the stability of Raman beam paths. Since the two addressing beams have independent beam paths, any mechanical perturbation along the beam paths causes fluctuations in $\phi_{\text{beam}}$. We perform Ramsey interferometry by driving optical-phase-sensitive \cite{InlekPRA2014} carrier transitions with two counter-propagating Raman beams to measure the laser coherence time, as shown in Fig. 4. This coherence time represents the timescale over which our Raman beam paths are interferometrically stable, and agrees with the 2-Hz drift used in the random walk simulation of $\phi_{\text{beam}}$.

\end{document}